\documentclass[journal]{IEEEtran}
\usepackage{amscd,graphicx,latexsym,multicol,cite,ulem,epsf}
\usepackage{verbatim}
\usepackage{graphics}
\usepackage{amssymb}
\usepackage[cmex10]{amsmath}
\usepackage{dsfont}
\usepackage{bbm}
\usepackage{setspace}
\usepackage{enumerate}
\usepackage{accents}
\newcommand{\qeq}{\accentset{(a)}{=}}

\usepackage[T1]{fontenc}
\usepackage[utf8]{inputenc}
\usepackage{soul}
\usepackage{float}

\begin{document}
\title{Partially-Distributed Resource Allocation in Small-Cell Networks}
\author{Sanam Sadr, \textit{Student Member, IEEE}, and Raviraj S. Adve, \textit{Senior Member, IEEE}\thanks{This work was sponsored by TELUS and the National Science and Engineering Research Council (NSERC) Canada. The authors are with the Edward S. Rogers Sr. Department of Electrical and Computer Engineering, University of Toronto, 10 King's College Road, Toronto, Ontario, Canada M5S 3G4 (email: \{ssadr, rsadve\}@comm.utoronto.ca).}}

\date{}
\maketitle

\begin{abstract}
We propose a four-stage hierarchical resource allocation scheme for the downlink of a large-scale small-cell network in the context of orthogonal frequency-division multiple access (OFDMA). Since interference limits the capabilities of such networks, resource allocation and interference management are crucial. However, obtaining the globally optimum resource allocation is exponentially complex and mathematically intractable. Here, we develop a \emph{partially} decentralized algorithm to obtain an effective solution. The three major advantages of our work are: 1) as opposed to a fixed resource allocation, we consider load demand at each access point (AP) when allocating spectrum; 2) to prevent overloaded APs, our scheme is dynamic in the sense that as the users move from one AP to the other, so do the allocated resources, if necessary, and such considerations generally result in huge computational complexity, which brings us to the third advantage: 3) we tackle complexity by introducing a hierarchical scheme comprising four phases: user association, load estimation, interference management via graph coloring, and scheduling. We provide mathematical analysis for the first three steps modeling the user and AP locations as Poisson point processes. Finally, we provide results of numerical simulations to illustrate the efficacy of our scheme.
\end{abstract}


\section{Introduction}
Each new generation of wireless communication systems promises to support a larger number of mobile users with higher data rates, new applications and ever-more stringent quality of service (QoS) requirements~\cite{kilper:11}. In a modern communication system where the mobile user is almost always connected to the network, supporting a large number of users with various applications results in a mix of traffic demands (bursty vs.~continuous, high vs.~low data rate) from the network point of view and high battery consumption from the user equipment (UE) point of view. Meeting these demands is getting harder largely due to the limited availability of transmission resources, most importantly wireless spectrum. This makes it impossible to completely separate concurrent transmissions in frequency, in turn, making interference the main factor that limits the capabilities of the wireless networks. Much of the relevant recent research focuses on the issue of interference e.g.,~\cite{ganti:08}, and network analysis in an interference-limited regime e.g.,~\cite{andrews:11,dhillon:11-2, jo:12, dhillon:12}.

In the cellular context, an effective approach to increase capacity is to reduce the distance between the transmitter and the receiver. This allows for a reduction in the transmit power and, in turn, improving the frequency reuse factor~\cite{andrews:08}. As an example of this approach, heterogeneous networks have been proposed to increase the area spectral efficiency\footnote{Area spectral efficiency is defined as the sum throughput per unit area that the system can provide per unit bandwidth.} and the overall capacity through a revised network topology. The idea is to introduce a multi-tier network where each tier, also referred to as a layer, differs mainly in the density of its access points (APs) and the AP transmit power, hence coverage area. The traditional macro-cellular network would be one layer in the heterogeneous network with carefully planned deployment, the lowest density and the highest transmit power; the small-cell layer, on the other hand, is characterized by essentially random deployment, a much higher AP density and low AP transmit power. The randomness in the AP locations in small cells and their significantly greater number within a chosen geographical area precludes globally optimized resource planning. This necessitates new analysis techniques and algorithms beyond those for the centrally planned macro-cellular layer. In this paper, we focus on resource allocation techniques in small cells and propose a \emph{partially}-distributed hierarchical scheme which can be applied to a \emph{large-scale} network.

\subsection{Related Work}
The analysis of the signal-to-noise-plus-interference-ratio (SINR) in a heterogeneous network has been presented for a single-layer~\cite{andrews:11} and multi-layer~\cite{dhillon:11-2, jo:12, dhillon:12} network using Poisson point processes (PPP). The presented analyses are based on the statistical distribution of the SINR in the network derived at a reference user randomly located in the cell. It is shown in~\cite{andrews:11} that in an interference-limited network, as is the case in small cells, the probability of coverage when the user is associated with a layer is independent of the AP density and the transmit power. The same result is shown in a multi-layer network with all the layers having the same layer association bias factor and path loss exponent~\cite{dhillon:11-2, jo:12, dhillon:12}. While the coverage analysis here is a measure of the level of the received SINR, it is only synonymous with a chosen QoS if the reference user is guaranteed some resources. In other words, in a network with high density of users, while the users might be in coverage when considering the received SINR, the rate available to the user depends on the scheduling algorithms used at each AP.

Several resource allocation schemes have been proposed in the literature to mitigate RF interference in small cells in the context of femtocell networks. Femtocells are essentially user-deployed, indoor, small cells. Regardless of the details of the system under consideration, they can be categorized into autonomous power control~\cite{barbieri:12, claussen:07, claussen:08, chandrasekhar2:09, ahmed:12} and adaptive spectrum allocation~\cite{chandrasekhar:08, perez:09, sousa:09, zhang:10,kim:09,zheng:11}. The main feature of the first group is to adjust the AP coverage by setting the transmit power high enough to service its users but low enough so as not to interfere with the other APs on the same frequency of operation. The schemes in the second group manage interference by ensuring orthogonality between interfering APs. 

A two-phase frequency assignment is proposed in~\cite{perez:09}, with a fixed, limited number of users per femtocell. Li et al.~\cite{sousa:09} viewed the user-deployed femtocells as the secondary system and the femtocell resource allocation as a cognitive spectrum reuse procedure. The idea is to adaptively adjust the channel reuse factor according to the location of the femtocell in the macrocell. Jointly optimizing power and spectrum, Kim et al.~\cite{kim:09} proposed a scheme to maximize the total system capacity in dense networks. Treating the macro-users as primary users, the authors in~\cite{li:12} prioritize them by performing a hand-over to the nearby femtocell whenever the small-cell interference is high. The graph-based approach proposed in \cite{zheng:11} maximizes the logarithmic average cell throughput to ensure proportional fairness among femtocells each serving a single user. A system level simulation of an open-access network was carried out by Claussen et al.~\cite{claussen:07,claussen:08}, and the obtained data rates at the reference users (one macro-user and one femto-user) were used to evaluate the system performance. Two main results are shown: 1) if autonomous power control is used by femtocells, adding APs has little impact on the macrocell throughput, and the impact is independent of the number of femtocells; 2) the total throughput significantly increases with the increase in the number of femtocell users, especially in the uplink. Similar results were reported in~\cite{dhillon:11-2, jo:12, dhillon:12}. When analyzing the system, it is assumed that either the reference user is guaranteed some resources, e.g., in~\cite{ganti:08, andrews:11, dhillon:11-2, jo:12, dhillon:12}, or only voice is considered, e.g., in~\cite{zhang:10}.

A more realistic simulation-based study of small-cell deployment in a heterogeneous network was reported by Coletti et al.~\cite{coletti:12}. The results suggest either coordination among layers or orthogonal spectrum allocation to improve outage rate. The authors of~\cite{mahmud:13} propose a combination of fractional frequency reuse (FFR) and orthogonal spectrum allocation in a two-tier network differentiating between commercial and home-based femtocells.

An ambitious goal in dense networks is to achieve optimal but decentralized resource allocation. The problem of decentralized power allocation was first addressed by Foschini et al.~\cite{foschini:93}. They showed that there exists a fully distributed algorithm which requires only local information if there exists a common, known, SINR at which the system performance is globally optimum, and there exists a feasible but unknown power vector that achieves this SINR. Unfortunately, these assumptions are hard to satisfy in practice~\cite{mitra}. The proposed distributed algorithms in~\cite{kiani:06,kiani:07} maximize the total system capacity ignoring user rate requirements and fairness among the users both within and among cells while~\cite{ahmed:12} aims for proportional fairness ignoring individual user rate requirements. To obtain a distributed solution, the authors in~\cite{kiani:06,kiani:07} simplify the network model to an ``interference-ideal" network where the total interference is constant and independent of user location in the cell.

\subsection{Our Approach and Contributions}
While there are several works on resource allocation in small-cell networks, as the literature survey above shows, it is hard to scale these algorithms to \emph{large-scale} networks with multiple hundreds of nodes. Specifically, we consider the downlink of a large-scale network of small cells. Instead of focusing on SINR, we attempt to provide users with their desired data rate as they declare to their corresponding AP. Each AP transmits at its full transmit power and we focus on frequency allocation to avoid interference. To maintain fairness among users, we formulate the resource allocation in the form of a max-min normalized rate problem, maximizing the minimum ratio between the achieved and the desired rates. This problem is, in general, NP-hard.

The main contribution of this paper is an effective solution to this resource allocation problem with reasonable computational complexity. We propose a hierarchical scheme by decomposing the problem into four steps. This scheme has several advantages. As opposed to a fixed spectrum allocation as in~\cite{chandrasekhar:08} and~\cite{zhang:10}, it considers the AP load, in terms of the number of users and their rate requirements, when allocating spectrum to the APs. This adaptivity in spectrum allocation allows for resources to follow user demands, i.e., high-rate users can be satisfied by a single AP. The proposed scheme is partially-distributed in the sense that three of the four steps are carried out locally and concurrently at each AP and, at worst, involve solving convex optimization problems using local information only. A single, graph coloring step must be executed at a central server. In contrast to the globally optimal solution requiring exponential complexity and global knowledge of channel state information, our hierarchical scheme imposes limited complexity and requires local knowledge only. 

Given the difficulty with applying available algorithms to large-scale networks, we compare the results of our scheme with a network with fixed number of channels allocated to each AP, and show how load-awareness can effectively reduce the outage rate. As far as possible, we provide a mathematical analysis for the proposed scheme based on stochastic geometry~\cite{baccelli:97}. In particular, we use independent homogeneous Poisson point processes for AP and user locations. PPPs have been shown to capture the inherent randomness in user and AP locations and yet provide tractable analyses compared to the grid-based models~\cite{andrews:11}. Necessarily requiring a number of simplifying assumptions, the analysis does provide results that track the related simulations and can, therefore, be used for system design.

The paper is structured as follows: in Section~\ref{system}, we describe the system model and formulate the resource allocation problem. The proposed hierarchical algorithm is presented in Section~\ref{sec:proposed} with the outage analysis based on homogeneous PPP. This section also presents a complexity analysis. Section~\ref{results} presents the simulation results in two parts: (a) comparing the performance of the proposed algorithm with its associated analysis; (b) comparing the performance with a fixed resource allocation scheme. The parameters used in the simulations are based on the Long Term Evolution (LTE) standard. Section~\ref{conclusion} concludes the paper with a discussion of the contributions.

\section{Downlink System Model and the Problem Statement} \label{system}
\begin{figure}[h!]
\center
\includegraphics [scale = 0.3]{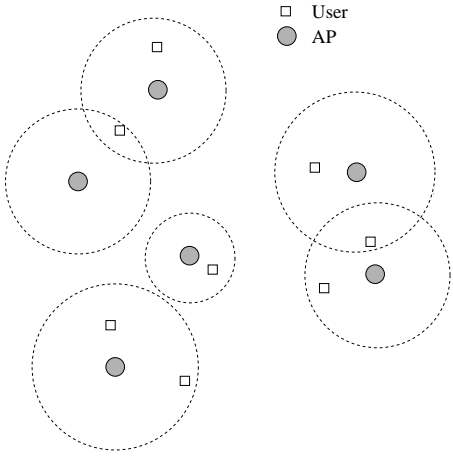}
\caption{Random distribution of APs and users in the network.}
\label{fig:femtonet}
\end{figure}

Figure~\ref{fig:femtonet} illustrates the network under consideration. The model comprises $K$ users and $L$ APs distributed randomly in the network. The macro base station (BS) is provided an orthogonal frequency allocation, and our analysis considers only users connecting to the small cells. If the BS allocation is not orthogonal, it can be easily incorporated into the algorithm as another transmitting node in the network with its own power budget. Associated with each AP is its potential coverage area which depends on the environment and the transmit power. Due to the random geographical distribution of access points, the coverage area for some (if using the same time-frequency resource) may overlap. In other words, some users might be in a location covered by more than one AP and transmissions from these APs would interfere.

The optimization problem is formulated in the context of OFDMA as in the LTE standard. There are $N$ effective frequency subchannels - physical resource blocks (PRBs) in LTE - available in the system each with a bandwidth of $B$. The channels between APs and users are modeled as frequency-selective Rayleigh fading with average power determined by distance attenuation and large scale fading statistics. The goal is to provide each user with its requested data rate. However, to achieve overall fairness in doing so, we formulate the problem to maximize the minimum normalized rate, i.e., max-min over all the users' achieved rates normalized by their requested data rates. The rate achieved on a specific channel is assumed to be given by its Shannon capacity; a gap function can be added to account for practical modulation and coding~\cite{goldsmith:97}. Each AP schedules its users in a manner to cancel the intra-cell interference. Therefore, the interference experienced by a user is due to the transmissions from all the access points, other than its own serving AP, that transmit on the same frequencies.

Under this setting, the general form of the resource allocation problem in the downlink is given by:
\begin{eqnarray}
\label{prob1}
\max_{\{p_{k,n}^{(l)}\}, \{S_l\}} \min_{k} & & \hspace*{-0.2in} \frac{1}{R_k}\left[\sum_{n = 1}^{N}B
    \log_{2}\left ( 1 +  \frac{p_{k,n}^{(l)} h_{k,n}^{(l)}}
    {\sum_{i=1, i \neq l}^{L}p_{k,n}^{(i)} h_{k,n}^{(i)} + \sigma^{2}} \right )\right], \nonumber \\
\mbox{subject to:} & & \hspace*{-0.2in} \sum_{n =1}^{N}\sum_{k \in S_{l}}p_{k,n}^{(l)} \leq P_{tot}, \; \; \;  l=1,2, ..., L, \nonumber\\
& & \hspace*{-0.2in} p_{k,n}^{(l)} \geq 0, \; \; \forall k,n,l, \label{eq:GlobalOpt} \\
& & \hspace*{-0.2in}S_{i} \cap S_{j} =  \emptyset \; \; i \neq j, \nonumber \\
& & \hspace*{-0.2in} \bigcup_{l =1}^{L} |S_{l}| = K,  \nonumber
\end{eqnarray}
where $h_{k,n}^{(l)}$ and $p_{k,n}^{(l)}$ are, respectively, the channel power gain and the transmit power from AP $l$ to user $k$ on subchannel $n$. $S_{l}$ is the set of users connected to and being serviced by AP $l$. $R_k$ is the required rate by user $k$ in bits per second (bps), $\sigma^{2}$ is the noise power, and $B$ is the bandwidth of each subchannel. The sum, $\sum_{i=1, i \neq l}^{L}p_{k,n}^{(i)} h_{k,n}^{(i)}$, is the cumulative interference experienced by user $k$ on subchannel $n$ from all the access points except the serving AP indexed by $l$. The first constraint is on the total transmit power of each AP, while the second ensures non-negative transmit powers on each subchannel. The third constraint ensures that the sets, $\left\{S_{l}\right\}_{l = 1}^L$, are disjoint, since each user is serviced by one and only one AP. The final constraint ensures that all the users in the system are scheduled by an access point. The sets, $\{S_l\}_{l=1}^{L}$, therefore, form a partition on the set of all users.

The objective of~\eqref{eq:GlobalOpt} is to find the optimal user associations, $\{S_{l}\}_{l=1}^{L}$, and power levels, $\{p_{k,n}^{(l)}\}_{l=1}^{L}$ determining which user should receive service from which AP on which subchannel, and how much power should be allocated to each subchannel. Being combinatorial, since it includes set selection, finding the optimal solution is exponentially complex. It seems infeasible from another point of view as well: it requires the knowledge of all the subchannels for all the users from all the APs at the central location. Getting this information to a central server would impose a huge overhead. Furthermore, this information needs to be updated every time the channel estimation is performed. Essentially, a resource allocation scheme based on global and perfect knowledge of SINR in a network of such scale is practically infeasible. This motivates developing partially-distributed, if suboptimal, solutions.

\section{Proposed Hierarchical Algorithm}\label{sec:proposed}
In response to the infeasibility of obtaining the globally optimum solution, we propose a partially-distributed resource allocation scheme to decompose the problem into four steps:
\begin{enumerate}
\item Cell association: each user is associated with the AP that offers the highest long-term \emph{average} received power (based, e.g., on a pilot and large-scale fading);
\item Load estimation: the load imposed by the users is \emph{estimated} by each AP based on its users' data rate requirements and \emph{average} channel gains;
\item Channel allocation: specific subchannels are allocated to APs based on coloring an interference graph;
\item Scheduling: each AP schedules its own users considering the users' required data rates and their \emph{instantaneous} channel gains.
\end{enumerate}
For each step, we derive the statistical distribution of the important quantities as accurately as possible, which are then used in analysis of the system performance, specifically the outage rate.

\subsection{Partially-Distributed Resource Allocation}
\subsection*{Step 1: Cell Association}
\label{proposed}
The cell association is based on the large scale fading only. This implies that the cell association is the same whether the system considers the uplink or the downlink. At each user, the received power from all the APs are measured, and the user is associated with the AP that offers the highest long-term average received power. This ensures maximum data rate on average and makes the cell association independent of the instantaneous channel gains. This cell association is consistent with the downlink model for the system analysis based on PPP considered in~\cite{andrews:11} and~\cite{jo:12}.

\emph{Analysis}: Considering only distance attenuation, each user's serving AP is the closest access point to the user. The statistical analysis of the connection distance is straightforward. Taking the user's location as the reference point, the connection distance is $d$ if there is no other access point closer to the reference point. Modeling the AP locations by a homogeneous Poisson point process with density $\lambda_{f}$, the cumulative density function (CDF) of $d$ can be written as:
\begin{equation}
\begin{array}{ll}
\label{shortest}
F_{D}(d)& = \mathds{P}(D < d) = 1 - \mathds{P}(N_{f}(A_{d}) = 0) \\
& \qeq 1 -  \exp(-\lambda_{f} \pi d^{2}). \\
\Rightarrow f_{D}(d) & = 2 \pi d \lambda_{f} \exp(-\lambda_{f}
\pi d^{2}), \hspace*{0.5in} d \geq 0,
\end{array}
\end{equation}
where $A_{d}$ is the area of a circle with radius $d$ and the user at the center, $\mathds{P}(B)$ denotes the probability of event $B$, and $N_{f}(A_{d})$ is the number of APs in $A_{d}$. $(a)$ results from the null probability of a 2-D Poisson process. Differentiating $F_{D}(d)$ with respect to $d$ gives the probability density function (PDF) in the final equation.

\subsection*{Step 2: Load Estimation}
With users having different rate demands, the objective at each AP is to estimate the minimum number of subchannels required to service its users. Each AP is aware of the requested rates and \emph{instantaneous} channel gains for all the users that it serves. However, it does not know which subchannels it will be allocated, and fading is frequency selective. Therefore, it estimates its load using only the \emph{average} channel gains. We emphasize that the load is defined here as the minimum frequency resources needed to meet the users' rate demands. We formulate this problem as a convex optimization problem \emph{at each AP} indexed by $l = 1, \dots, L$ given by:
\begin{eqnarray}
\label{prob2}
\min_{n_{k}, P_k}& & \sum_{k \in S_{l}} n_{k}, \nonumber \\
\mbox{subject to:} & &  n_{k}B\log_{2}\left ( 1 +
    \frac{P_{k} H_{k}}{n_{k}\sigma^{2}} \right) \geq R_{k}, \; \forall k \in S_{l},  \nonumber \\
&& \sum_{k \in S_{l}}P_{k} \leq P_{tot},\\
&& P_{k} \geq 0, n_{k} \geq 0, \;\; \forall k \in S_{l}. \nonumber
\end{eqnarray}
Here, $n_k$ is the number (can be a fraction) of subchannels that AP $l$
budgets for user $k \in S_l$. As before, $S_l$ is the set of users supported by AP $l$. $P_{k}$ and $H_{k}$ are, respectively, the total power allocated to and the \emph{average channel power} seen by user $k$. The first constraint ensures that the access point requests adequate resources to meet its users' demands. The objective is to minimize the total amount of spectrum needed by AP $l$. This is important since it affects the density of the interference graph in the next step.

\emph{Analysis}: The load of the $l$-th AP in terms of the required amount of spectrum is a random variable given by:
\begin{equation}
\label{AP_load1}
N_l = \sum_{k = 1}^{m}n_{k},
\end{equation}
where $m=|S_l|$ is itself a random variable representing the number of users connected to AP $l$. The CDF of $N_l$ in the general form is given by:
\begin{equation}
F_{N_{l}}(n_{l}) = \mathds{E}_{m, \{n_{k}\}}\left [ \mathds{P} \left( \sum_{k=1}^{m} n_{k}\leq n_{l}\right) \right ],
\end{equation}
where $\mathds{E}_{m,\{ n_{k}\}}$ denotes expectation with respect to $m$ and $\{n_{k}\}_{k=1}^{m}$, and $n_{k}, k=1,\dots,m$ are i.i.d~random variables representing the required spectrum of each user. A thorough mathematical analysis of the load requires the knowledge of the number of users in each AP's coverage area, also referred to as \emph{Voronoi} cells, and the users' distance to the serving AP. Due to the intractability of the problem, we derive the CDF for a special case where (i) all the APs have the same coverage area \emph{on average}, (ii) equal power is allocated to each subchannel to estimate the load, and (iii) all the users connect at a distance equal to the most probable distance from the AP. We now state and prove the main result at this step.

\emph{Proposition}: Under the assumptions stated above, the CDF of the AP load is given by:
\begin{equation}
F_{N_{l}}(n_{l}) = \mathcal{B}(n_l/n^*,K,p),
\end{equation}
where $\mathcal{B}(n_l/n^*,K,p)$ is the CDF of a binomial distribution with $K$ trials and probability of success $p = 1/L$ evaluated at $n_l/n^*$ ($n^*$ is defined later in~\eqref{approx} to be the number of subchannels required by each user).

\begin{proof}
In a network with $L$ APs, the probability that a user is in the coverage area of a specific AP is $1/L$. Since user locations are random and independent, the probability mass function of the number of users $m$ connecting to an access point is a binomial distribution given by:
\begin{equation}
\label{eq:binomial}
P_{u}(m) = {K \choose m} p^{m}q^{K-m},
\end{equation}
where $p = 1/L$ and $q = 1 - p$.

From the first constraint in \eqref{prob2}, the minimum required number of subchannels for user $k$ with equal transmit power on each subchannel can be rewritten as:
\begin{equation}
\label{prob3}
\displaystyle n_{k} = \frac{R_{k}}{B\log_{2}\left (1 + P_{tot}H_{k}/N\sigma^{2} \right)},
\end{equation}
where $H_{k} = L_{0}d^{-\alpha}$, $d$ is the distance between the AP and the user, $L_{0}$ is the path loss at the reference distance ($r_{0} = 1$m), and $\alpha$ is the path loss exponent. Setting $\gamma_{0} = P_{tot}L_{0}/N\sigma^{2}$ and using $F_{D}(d)$ derived in \eqref{shortest}, the CDF of the user load $F_{n_{k}}(n)$ can be written as:
\begin{equation}
\label{Nu_cdf}
\begin{array}{ll}
F_{n_{k}}(n)  & =  \mathds{P}(n_{k}< n) \\
 & = \displaystyle \mathds{P}\left( \frac{R_{k}}{B \log_{2}\left( 1 + \gamma_{0} d^{-\alpha} \right)}  < n \right) \\
 & = \displaystyle \mathds{P}\left( \log_{2} (1 + \gamma_{0} d^{-\alpha}) > \frac{R_{k}}{nB} \right) \\
 & = \displaystyle \mathds{P}\left( d < \left ( \frac{2 ^{R_{k}/nB} - 1}{\gamma_{0}} \right)^{-1/\alpha} \right) \\
& = \displaystyle F_{D}\left( \left ( \frac{2 ^{R_{k}/nB} - 1}{\gamma_{0}} \right)^{-1/\alpha} \right).
\end{array}
\end{equation}
 As shown in Fig.~\ref{fig:Nu_cdf} in Section \ref{results}, this CDF approaches a step function as $\lambda_{f}$ increases. In other words, in a dense network of small cells, the variance of the user load decreases. This motivates one to represent the user load in high-density AP networks with a single number rather than a random variable. Amongst various possibilities, the \emph{most probable link distance} best predicts this number.

Differentiating $f_{D}(d)$ with respect to $d$ and setting it to zero, the most probable link distance is given by:
\begin{equation}
\label{linkD}
 d^{*} = \sqrt\frac{1}{2\pi \lambda_{f}}.
\end{equation}
Inserting $d^{*}$ in \eqref{prob3}, the number of subchannels required by each user is given
by:
\begin{equation}
\label{approx}
 n^{*} = \frac{R_{k}}{B \log_{2}\left ( 1 + \gamma_{0}(d^{*})^{-\alpha} \right)}.
\end{equation}
Using \eqref{AP_load1}, the CDF of the AP load is then derived as:
\begin{equation}
\label{APLoad}
\begin{array}{ll}
F_{N_{l}}(n_{l}) & = \mathds{P}( N_{l} \leq n_{l}) = \mathds{P}(n^{*}m \leq n_{l}) \\
& = \mathds{P}(m \leq n_{l}/n^{*}) \\
& = \mathcal{B}(n_l/n^*,K,p),
\end{array}
\end{equation}
where the last equation results from the distribution of $m$ derived in \eqref{eq:binomial} and the proof is complete. 
\end{proof}

For a large $K$, the binomial distribution is very well approximated by a Poisson distribution with parameter $\eta = K/L$ when $\eta$ is small, and by a normal distribution when $\eta$ is large~\cite{Leon-Garcia}. In this paper, we will make use of the Poisson approximation.

\subsection*{Step 3: Channel Allocation Among APs Using Graph Coloring}
After steps 1 and 2, users have been assigned to APs and the APs have estimated their loads. We now come to the crucial step of allocating subchannels to APs. Specifically, the objective at this step is to allocate the spectrum to the APs considering their load and the interference they can potentially cause to other small cells. In this paper, we consider a resource allocation scheme which avoids interference. To do so, we ensure that two neighboring small cells do not use the same frequencies. Small cells are considered neighbors if they potentially interfere with each other, i.e., their potential coverage areas overlap. Unlike the previous two steps (and the next step), this allocation is \emph{centralized}.

All the access points report their load $\left\{N_l\right\}_{l=1}^L$ to the central server. Ideally, the central server should allocate an orthogonal set of subchannels to every AP that also meets its users' requirements. Given $N$ total available subchannels, if $N\geq \sum_{l=1}^L N_l$, each AP is easily satisfied. Realistically, however, this is highly unlikely; hence, the server must reuse channels across multiple APs. This can cause interference, and so the allocation must ensure that the interfering APs (APs with overlapping coverage areas) are assigned orthogonal frequency resources. As a consequence, it may be that all APs' load demands cannot be satisfied. Alternatively, the goal is to assign subchannels to APs \emph{proportional} to their estimated load while eliminating the interference among them. To do so, we use graph coloring by the central server.

Graph algorithms have been used as a tool for channel assignment in multi-cellular networks, e.g., in~\cite{narayanan:02, chang:08, chang:09, Gao:11, kim:11,patero:13}, with the nodes representing either access points or users. Chang et al.~\cite{chang:08} formulated the spectrum allocation in a macrocellular network in the form of max $\mathcal{K}$-Cut with a fixed number of channels (or colors). Each node in the graph corresponds to a mobile device or user. The interference among users is denoted by weighted edges taking into account not only the distance between the users but also the anchor (serving) and the neighboring base stations. The objective is to partition the users into $\mathcal{K}$ clusters with maximum inter-cluster weight. This technique allows for asymmetrical channel allocations among the base stations. Authors in~\cite{Gao:11} proposed a two-step graph coloring approach for multicell OFDMA networks in which the users are clustered in a manner to minimize the total number of colors based on geographic user locations. In the second step, the subchannels are allocated based on instantaneous channel conditions. In graph-based schemes, wherever users correspond to graph nodes as in~\cite{Gao:11} and~\cite{patero:13}, user mobility results in rapid changes of the interference graph. Since we deal with small cells in a dense network, this computation is added to the signalling overhead due to hand-off and synchronization among APs making it impractical. The authors in~\cite{chang:09} differentiate between the cell centre and cell edge hence allowing for FFR. This approach assumes large cells, an assumption that is not valid here. Finally, the two-step spectrum allocation algorithm proposed in~\cite{kim:11} uses the instantaneous channel information in deriving femtocells' utilities while coloring the graph resulting in increased complexity and signalling overhead.

\begin{figure}
\center
\includegraphics [scale = 0.23]{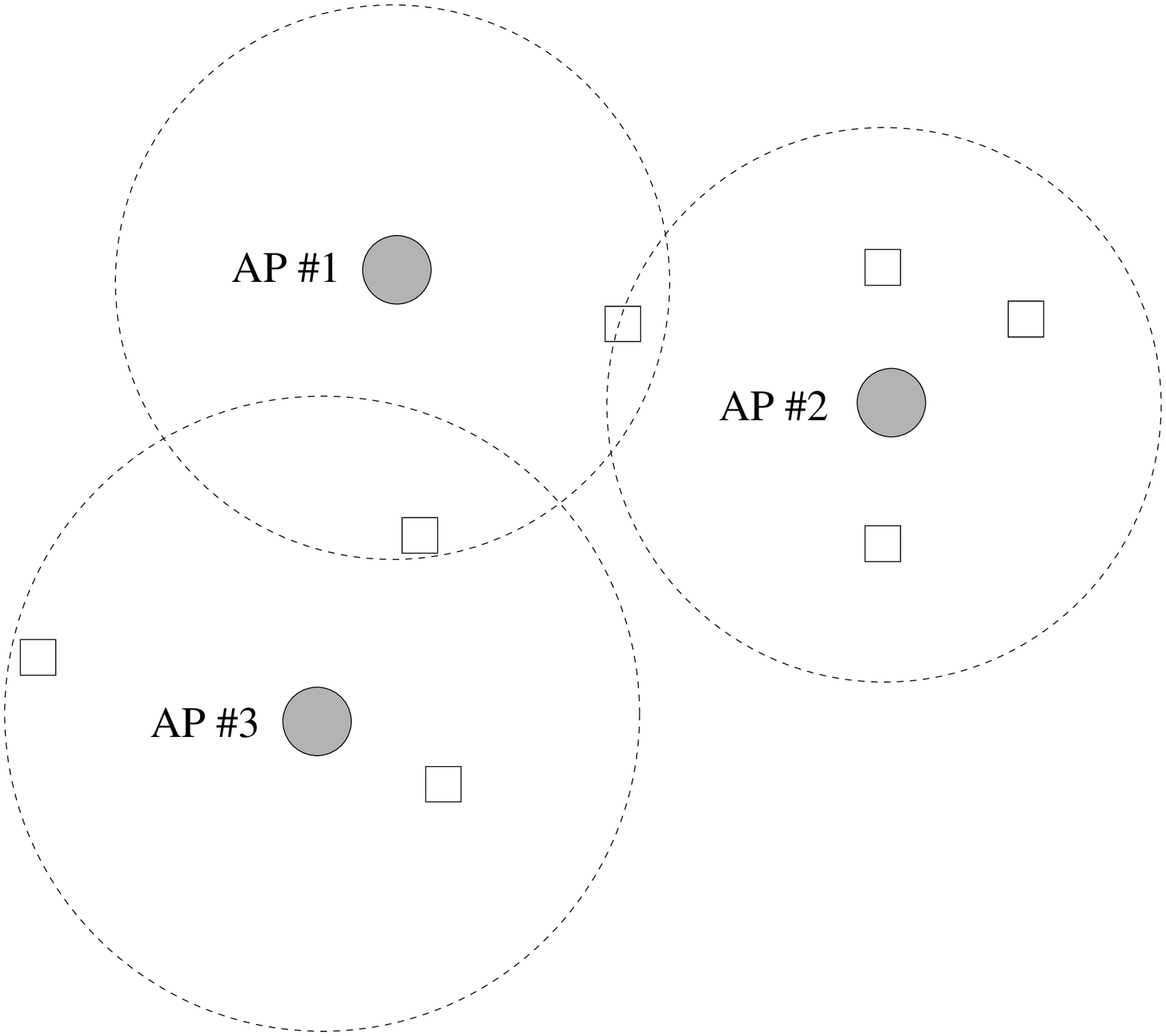}
\caption{A network of 3 APs. $N_1 = 1, N_2 = N_3 = 3$.}
\label{fig:interference}
\center
\includegraphics [scale = 0.23]{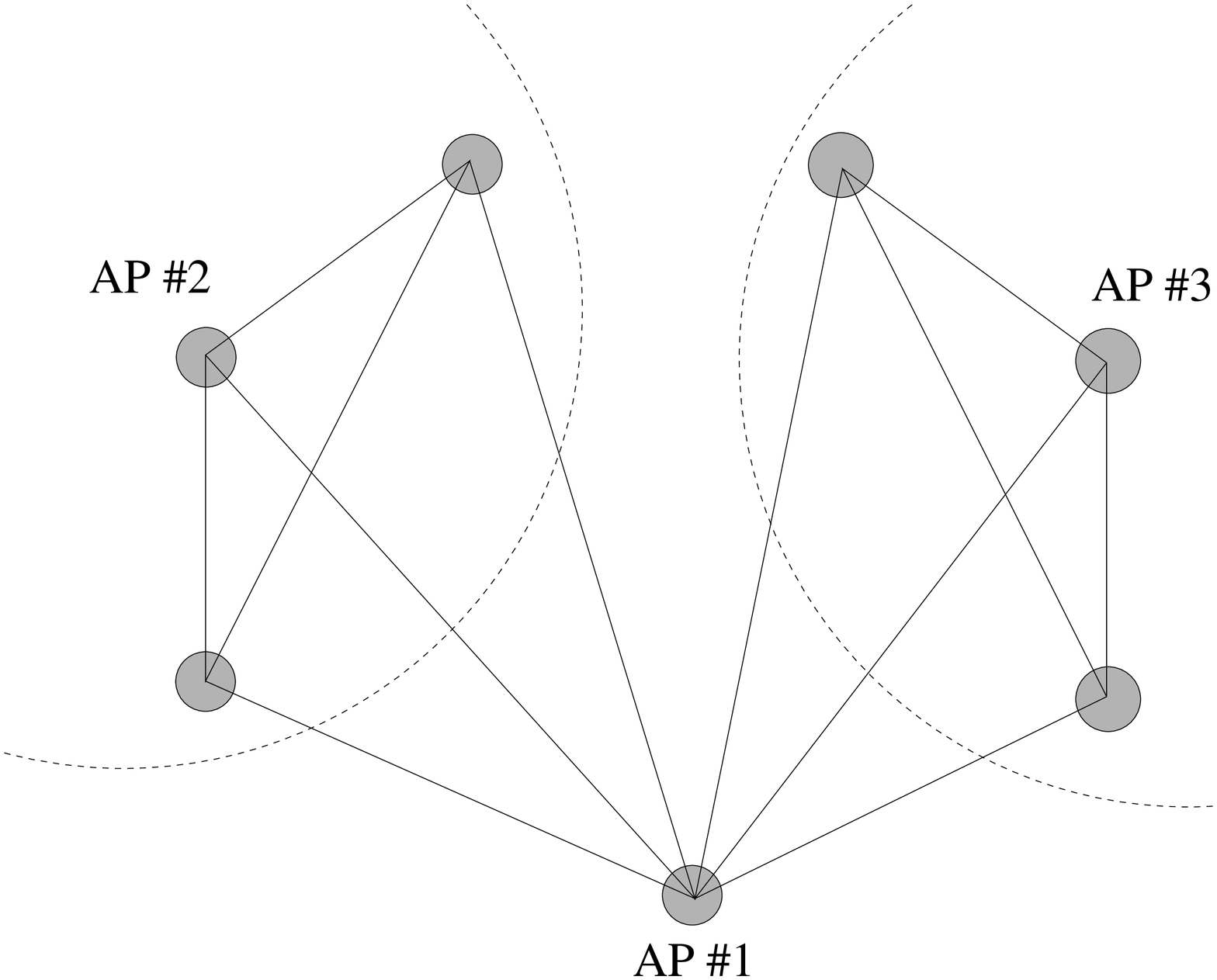}
\caption{Interference graph corresponding to Fig.~\ref{fig:interference}\label{fig:inter_graph}.}
\center
\includegraphics [scale = 0.4]{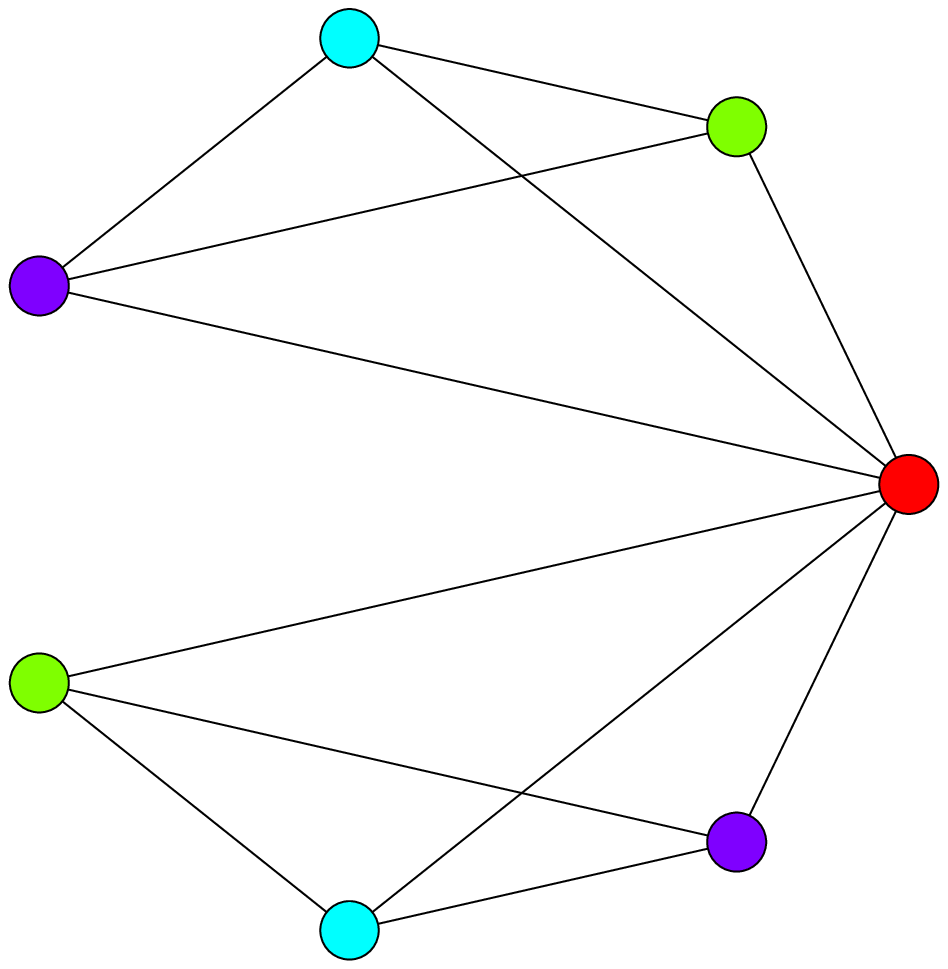}
\caption{Graph coloring corresponding to Fig.~\ref{fig:inter_graph}.
Minimum number of colors is 4, with both optimal and suboptimal coloring algorithms. }
\label{fig:coloring}
\end{figure}

In our approach, the nodes of the graph represent access points. An edge connects two nodes if they potentially interfere based on \emph{large-scale} statistics. We make this choice to ensure that the graph does not change rapidly with each channel realization. While here we use an unweighted graph, this is not fundamental to the proposed scheme. A weighted graph can very well be used instead, at the cost of increased complexity as long as the edge weights correctly reflect the intensity of the interference between any two nodes. Since each color corresponds to a single subchannel, to account for the AP loads, we modify the interference graph as follows: as opposed to the conventional approach, AP $l$ is represented by not one but $\lceil N_{l}\rceil$ nodes forming a complete subgraph ($\lceil\cdot\rceil$ denotes the ``ceiling" function). The problem of channel assignment among APs becomes a graph coloring problem where two interfering nodes (nodes connected with an edge) should not be assigned the same color. An example of a three AP network with (estimated) $N_1 = 1, N_2 = N_3 = 3$ is illustrated in Fig.~\ref{fig:interference}. The corresponding interference graph is shown in Fig.~\ref{fig:inter_graph}. AP \#1 potentially interferes with AP \#2 and AP \#3. So, allocations to AP \#1 cannot be reused for AP \#s 2 or 3. However, since AP \#2 does not interfere with AP \#3, frequencies can be reused across these two APs.

One solution to the coloring problem is illustrated in Fig.~\ref{fig:coloring}. It is worth noting that the coloring is not unique. For example, a simple index shift (a re-ordering of the association between the graph and the frequency slots) is an equally valid solution to the graph coloring problem. Amongst these many solutions, there is one optimal solution that best meets the demands of the individual APs based on the specific instantaneous realizations of user-AP channels. However, none of this information is available at the central sever; this lack of optimality is the penalty for using a distributed algorithm with limited knowledge of CSI.

For arbitrary graphs, graph coloring is an NP-hard problem. The optimal coloring is possible with low complexity algorithms if the interference graph is sparse such that each node is connected to at most $N$ nodes where $N$ is the total number of available colors. Such graphs can be colored with a modified Breadth First Search (BFS) algorithm with complexity of $\mathcal{O}(|V| + |E|)$ with $|E| = \mathcal{O}(|V|)$ where $|E|$ and $|V|$ are the cardinality of edges and vertices respectively. We adopt the heuristic (greedy) algorithm proposed by Br\'{e}laz~\cite{Brelaz:79}: at every iteration, the vertex which is adjacent to the greatest number of differentely-colored neighbours is colored, with a new color if necessary (until colors are exhausted). A major advantage of our proposed hierarchical scheme is that by carrying out the graph coloring step in a distributed manner as proposed in~\cite{Oleg}, we achieve a fully distributed scheme.

\emph{Outage Analysis}: While coloring an interference graph results in a higher spatial reuse factor for a channel, outage is inevitable if an access point and its interfering neighbours require more than the available number of subchannels. Let $\tilde{p}$ be the AP pilot power and $\tau$ be its receive threshold determining the coverage area. Two APs separated by $\tilde{r} = 2\tilde{d}$ interfere if the received power at the midpoint of the distance between the two satisfies $\tilde{p}\tilde{d}^{-\alpha} > \tau$. Now, taking a randomly chosen AP as the reference, we prove the following statement.

\emph{Proposition}: The probability that the reference AP and its neighbours will not have enough spectrum, hence leading to an outage, is given by:
\begin{equation}
P_{o} = 1 - \sum_{\tilde{L} = 1}^{L} \left [ \mathit{Pois}(N/n^{*}, \tilde{L}\eta)\mathds{P}\left ( N_{f} (A_{\tilde{r}}) = \tilde{L} \right ) \right ],
\end{equation}
where $\tilde{L}$ is the number of APs in the area $A_{\tilde{r}} = \pi \tilde{r}^{2}$, 
$\mathds{P}\left(N_{f}(A_{\tilde{r}}) = \tilde{L}\right) = \frac{e^{-\lambda_{f}A_{\tilde{r}}}}{\tilde{L}!}(\lambda_{f}A_{\tilde{r}})^{\tilde{L}}$, and $\mathit{Pois}(N/n^{*}, \tilde{L}\eta)$ is the CDF of the Poisson random variable with mean $\tilde{L}\eta$ evaluated at $N/n^{*}$.

\begin{proof} 
Let $\tilde{L}-1$ be the total number of APs interfering with the reference AP, i.e., $\tilde{L}$ APs cannot share the same subchannel. The CDF of the minimum required number of subchannels $\tilde{N}$ such that our reference AP can support its users (without interfering with its neighbours) is given by:
\begin{equation}\label{systemLoad}
\begin{array}{ll}
F_{\tilde{N}}(\tilde{n}|\tilde{L}) & = \mathds{P}(\tilde{N} \leq \tilde{n}|\tilde{L}) \\
& = \mathds{P}\left ( \sum_{l = 1}^{\tilde{L}} N_{l} \leq \tilde{n}|\tilde{L} \right) \\
& = \mathit{Pois}(\tilde{n}/n^{*}, \tilde{L}\eta),
\end{array}
\end{equation}
where we use the fact that the sum of $\tilde{L}$ independent Poisson random variables with mean $\eta$ is a Poisson random variable with mean $\tilde{L}\eta$. The probability of outage is then given by:
\begin{equation}
\begin{array}{ll}
P_{o} & =  \mathds{E}_{\tilde{L}} \left [ \mathds{P}(\tilde{N} > N) \right ] = \mathds{E}_{\tilde{L}}[1 - F_{\tilde{N}}(N)]  \\
& = \displaystyle \sum_{\tilde{L} = 1}^{L} \left [1 - F_{\tilde{N}}(N|\tilde{L}) \right ]\mathds{P}\left ( N_{f} (A_{\tilde{r}}) = \tilde{L} \right ) \\
& = \displaystyle 1 - \sum_{\tilde{L} = 1}^{L} \left [ \mathit{Pois}(N/n^{*}, \tilde{L}\eta)\mathds{P}\left ( N_{f} (A_{\tilde{r}}) = \tilde{L} \right ) \right ],
\end{array}
\end{equation}
and the proof is complete. 
\end{proof}
Unfortunately, it does not appear that this final expression can be further simplified. However, the summation is easily evaluated numerically. In a system with $L$ points of a Poisson process (representing APs) uniformly distributed, $\tilde{L}$ is binomial with $L$ trials and probability of success $(\tilde{r}/R_{c})^2$ where $R_{c}$ is the radius of the macrocell under consideration. $F_{\tilde{N}}(\tilde{n})$ is the CDF of a Poisson random variable and can be calculated using the \emph{incomplete gamma function}.

\subsection*{Step 4: Resource Allocation Among Users}
At the end of step 3, each AP is assigned an integer number of subchannels without interfering with its neighbours. The problem at each AP is now reduced to maximizing the minimum rate of the users it services relative to their requested data rate. In doing so, the AP considers the \emph{instantaneous} CSI of the subchannels it has been assigned. In this regard, it is worth restating that the previous three steps were based on \emph{average} channel powers.

Let $\bar{N}_l$ be the number of subchannels assigned to AP $l$ in Step 3; this is not necessarily equal to its estimated requirement $N_{l}$. The scheduling problem at each AP is formulated as:
\begin{eqnarray}
\label{prob4}
\max_{p_{k,n}, c_{k,n}} \min_{k} & & \frac{1}{R_k}\left[B\sum_{n = 1}^{\bar{N}_{l}}c_{k,n}
    \log_{2}\left ( 1 +  \frac{p_{k,n} h_{k,n}}{c_{k,n}\sigma^{2}} \right)\right]
    \nonumber \\
\mbox{subject to} & & \sum_{n =1}^{\bar{N}_l}\sum_{k \in S_{l}}p_{k,n} \leq P_{tot}, \hspace*{0.1in}
p_{k,n} \geq 0 \nonumber \\
& & \sum_{k \in S_{l}} c_{k,n} = 1, \nonumber \\
& & c_{k,n} \geq 0 \; \; \forall n,\; \; k \in S_{l}
\end{eqnarray}
where $c_{k,n}$ is the fraction of subchannel $n$ allocated to user $k$. $h_{k,n} $ and $p_{k,n}$ are the channel power gain and the transmit power to user $k$ on subchannel $n$. This is a standard convex optimization problem. An even simpler alternative is to divide power equally amongst the $\bar{N}_l$ subchannels, $p_{k,n} = P_{tot}/ \bar{N}$, leading to a linear program in which users share the resources using time-division\footnote{Standards such as LTE provide the ability to reassign physical resource blocks every millisecond. Such flexibility is reflected here as time-sharing of the subchannels.}.

\subsection{Complexity Analysis}
\paragraph*{Step 1 - cell association}
Each user connects to the AP with the highest average received power. Finding the AP with the maximum received power requires $L$ comparisons at each user. Hence, the complexity of this step is of the order $\mathcal{O}(L)$ for each of $K$ users.
\paragraph*{Step 2 - load estimation}
This is a convex optimization problem with the complexity depending on the solution method, e.g., the interior-point method or Newton-Raphson. Furthermore, the number of iterations in each depends on the stopping criterion. In Newton-Raphson method, the computational complexity mainly results from finding the update direction. It is shown that the computational complexity of each iteration is $\mathcal{O}(M^{3})$, where $M$ is the number of users connected to one AP. The details are provided in the Appendix.
\paragraph*{Step 3 - spectrum allocation among APs}
This step consists of two smaller steps. 1) Forming the interference graph: any two APs closer than a threshold distance are connected with an edge. Hence, the complexity of this step is of the order $\mathcal{O}(L^2)$. 2) Graph coloring: the complexity depends on the density of the graph algorithm as provided in Step 3 of Subsection~\ref{proposed}. Since this step is carried out at the central unit, with slower changes compared to locally solved problems, more sophisticated algorithms can be used.
\paragraph*{Step 4 - scheduling}
The normalized rate scheduling at this step is a modified version of the problem formulated by Rhee et al.~\cite{rhee:00}. A special case is when equal transmit power is used on all the subchannels leading to close to optimum performance when the system benefits from user-channel diversity. The proposed suboptimal subchannel allocation with equal transmit power has complexity of $\mathcal{O}(M*\bar{N_{l}})$, where $\bar{N_{l}}$ is the number of PRBs allocated to the AP.

\section{Simulation Results}\label{results}
In this section, we evaluate the performance of the proposed scheme and the mathematical analysis presented in the previous section. The simulations are based on the LTE standard closely following~\cite{forum}. The downlink transmission scheme for an LTE system is based on OFDMA where the available spectrum is divided into multiple subcarriers each with a bandwidth of 15kHz. Resources are allocated to users in blocks of 12 subcarriers referred to as physical resource blocks; hence, the bandwidth of each PRB is 180kHz and is used as the signal bandwidth in calculating the noise power. The receiver noise power spectral density is set to -174dBm/Hz with an additional noise figure of 9dB at the receiver. Here, we consider the maximum LTE bandwidth (20MHz); $N=50$ PRBs are allocated to the small-cell network. The APs are distributed within a circle of radius 100m, i.e., covering $3.14 \times 10^4$ m$^2$. Table~\ref{tab:SimulationParameters} lists the parameters used in all the simulations, unless otherwise specified.

\begin{table}[h]
\center
\caption{Simulation Parameters\label{tab:SimulationParameters}}
\begin{tabular}{| c | c |}
\hline
Parameters & Value \\
\hline \hline
Carrier frequency & 2 GHz  \\
Channel bandwidth & 20 MHz \\
Carrier spacing & 15 kHz \\
Resource block ($B$) & 180 kHz \\
Number of PRBs available ($N$) & 50 \\
\hline
Transmit power & 20dBm \\
Antenna gain & 0dB \\
Antenna configuration & 1 $\times$ 1\\
\hline
Noise Figure in UE & 9dB\\
\hline
Minimum distance of user & 1m from AP \\
Penetration loss & 10dB/3dB \\
(wall/window) & \\
\hline
$\tilde{d}$ & 20m \\ \hline
Region covered ($R_{c})$ & Circle of radius 100m \\ \hline
\end{tabular}
\end{table}

The path loss between the access point and the user accounts for indoor and outdoor propagation:
\begin{equation}
{\rm PL} = 38.46 + 20\log_{10}(d_{in}) + 37.6\log_{10} (d) + L_{p} + L_s
\end{equation}
where $d_{in}$ is the distance between the access point and the external wall or window and has a uniform distribution between 1m and 5m; $L_{p}$ is the penetration loss and is set to 10dB or 3dB (with equal probability) representing an external wall and window respectively; $L_s$ accounts for shadowing and is modeled by a log-normal random variable with standard deviation of 10dB. Finally, assuming Rayleigh fading, the instantaneous power of the received signal is modeled as an exponential random variable with the mean equal to the average received power~\cite{hansen:77}. In the mathematical analysis, the path loss exponent, $\alpha$, is set to 3. The multipath environment is such that the fading is effectively flat for the 12 subcarriers in one PRB but rich enough to yield an independent fade on each PRB. Each PRB is then allocated to a user for a \emph{subframe} duration of 1ms.

\subsection{Validating the Mathematical Analysis}
\begin{figure}
\centering
\includegraphics [scale = 0.35]{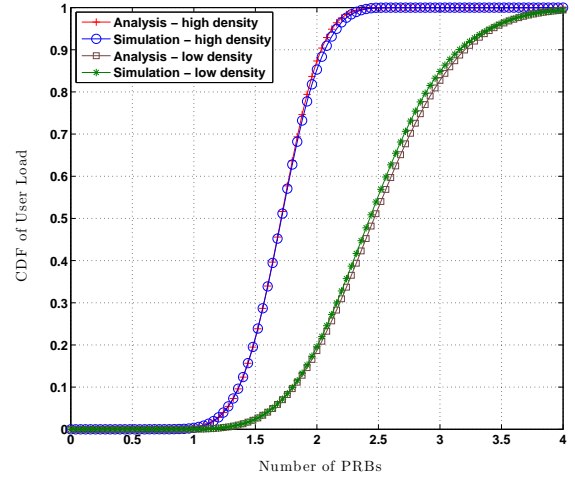}
\caption{CDF of the user load $F_{n_{k}}(\cdot)$ for high ($\lambda_{f} =1/100m^{2}$) and low ($\lambda_{f} = 1/1000m^{2}$) AP density. $R_{k} = 5$Mbps. }
\label{fig:Nu_cdf}
\end{figure}

First, we validate the mathematical analysis of the proposed hierarchical algorithm. Figure~\ref{fig:Nu_cdf} plots the CDF of the user load $F_{n_{k}}(\cdot)$ for two cases:  high ($\lambda_{f} =1/100$ or 1 AP per 100$m^2$ corresponding to 314 APs) and low ($\lambda_{f} = 1/1000$) AP density. At each run of the simulation, the APs and the users are randomly located in the cell according to the given densities with user density denoted by $\lambda_{u}$. Each user connects to the closest access point. The user load is then calculated for a randomly chosen user (as the reference) with equal transmit power on all the PRBs considering only the distance attenuation. The result is compared to the CDF derived in \eqref{Nu_cdf}. As is clear, the analysis matches the simulation results exactly. Also, as expected, the CDF is shifted to the right in a network with lower AP density; the reason is a larger connection distance and a higher load (measured in terms of required subchannels) imposed by the user for the same data rate. Crucially, the figure shows that the CDF of the individual user load approaches a step function at a higher AP density. This justifies the approximation in~\eqref{approx}.

\begin{figure}
\centering
\includegraphics [scale = 0.35]{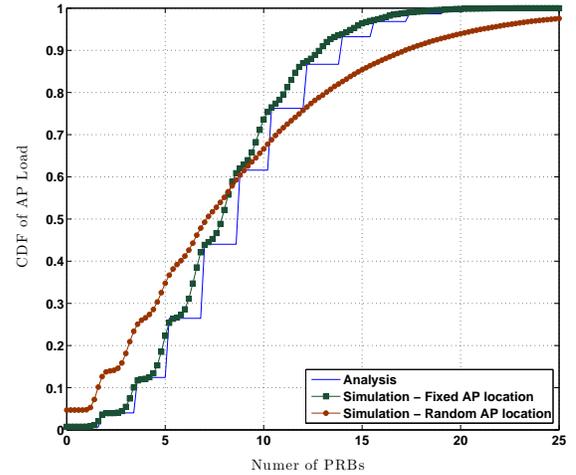}
\caption{CDF of the AP load $F_{N_{l}}(\cdot)$. $\lambda_{f} =1/(100m^{2})$, $\lambda_{u} = 5\lambda_{f}$ and $R_{k} = 5$Mbps.}
\label{fig:Nf_cdf2}
\end{figure}
\begin{figure}
\centering
\includegraphics [width=8cm,height=8cm,keepaspectratio]{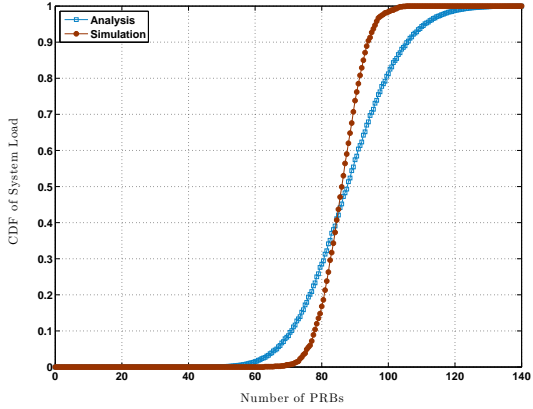}
\caption{CDF of the system load $F_{\tilde{N}}(\cdot)$. $\lambda_{f} =1/(100m^{2})$, $\lambda_{u} = 5 \lambda_{f}$ and $R_{k} = 1$Mbps.}
\label{fig:Ns_cdf}
\end{figure}

Figure~\ref{fig:Nf_cdf2} plots the CDF of the load of a randomly chosen access point in two scenarios: (i) random AP location as in PPP; (ii) fixed AP location with equal coverage area for all the APs. In both cases, the load of the reference AP is the sum of its users' load. The CDF derived in~\eqref{APLoad} slightly deviates from the simulation results due to two main simplifying assumptions: 1) all APs have the same coverage area on average leading to a binomial distribution for the number of users connecting to an AP; 2) the most probable connection distance has been assumed for all the users connecting to an AP based on the AP density in the system. In a system where APs are randomly located, although uniformly distributed, they might have different coverage areas. For the purpose of comparison, a network with fixed AP locations is considered with equal coverage area for all APs. In such a system, $P_{u}(m)$ follows the binomial distribution. A small deviation still exists due to the second simplifying assumption.

\begin{figure}[h]
\center
\includegraphics [width=8cm,height=8cm,keepaspectratio]{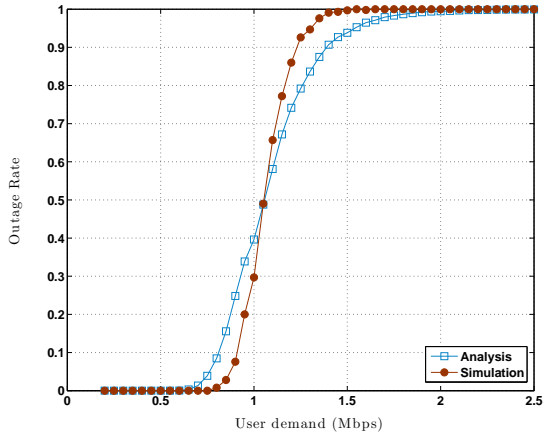}
\caption{Outage rate versus user demand. $\lambda_{f} =1/(200m^{2})$ and $\lambda_{u} = 5 \lambda_{f}$.}
\label{fig:OutageAnalysis}
\end{figure}

The CDF of the system load, $F_{\tilde{N}}(\cdot)$ is shown in Fig.~\ref{fig:Ns_cdf}. The system load is the sum of the total number of PRBs required by a randomly chosen access point (as the reference) and all its interfering APs derived in \eqref{systemLoad}. If the distance between two APs is no more than $2\tilde{d}$, where $\tilde{d}$ is the radius of the coverage circle of one AP, the pair are assumed to interfere with each other. A lower SNR threshold results in a larger coverage area for each AP and a denser interference graph. Note that, for any SNR threshold, this is the worst-case scenario assuming the user is in the midpoint of the distance between the two APs. In practice, whether two APs interfere can be estimated more accurately by each AP based on a pre-defined coalition threshold \cite{liu:12} and reported to the central server (most protocols allow an access point to keep a ``neighbour" list). Figure~\ref{fig:OutageAnalysis} plots the corresponding outage probability as a function of the common user demand $R_k = R$. Again, the analysis captures the essential behaviour of the outage probability with a slight mismatch due to the simplifying assumptions mentioned before for tractability.

\subsection{Performance Comparison}
In this section, we illustrate the performance of the proposed hierarchical scheme and compare its performance with a fixed-allocation scheme. The globally optimal solution through exhaustive search is impossible to obtain in a reasonable time due to its exponential complexity and so is not compared to.

The fixed-allocation scheme is as follows: each AP is assigned $N_{AP}$ PRBs randomly chosen out of the $N$ PRBs available to the small-cell network. The cell association and user level scheduling is the same for both algorithms. Hence, the main differences between the fixed-allocation and the proposed hierarchical scheme are the element of interference management and the effect of load estimation in dynamic distribution of PRBs among APs. The purpose of such dynamic distribution is to improve the user's achieved rate in the whole system proportional to its demand. A user is considered to be in outage when it receives less than its required data rate.

\begin{figure}[h]
\center
\includegraphics [width=8cm,height=8cm,keepaspectratio]{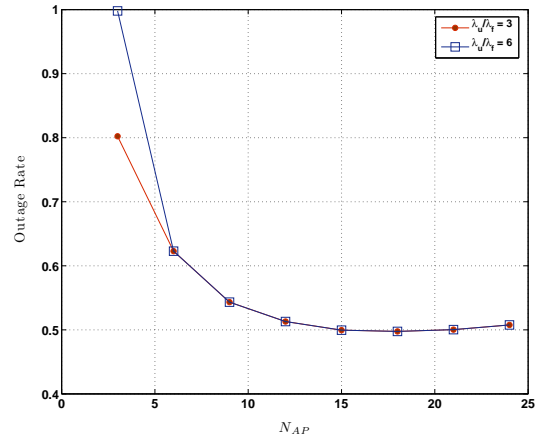}
\caption{Outage rate as a function of the fixed number of PRBs assigned to each AP. $\lambda_{f} = 1/(200m^{2})$, $\lambda_{u} = 3 \lambda_{f}$ and $\lambda_{u} = 6 \lambda_{f}$. $R_{k} = 1.5$Mbps.}
\label{fig:SFoutagePRB}
\end{figure}

In a network with fixed spectrum allocation, $N_{AP}$ affects the density of the interfering APs~\cite{andrews:11}. We first consider the performance of the fixed-allocation scheme as a function of $N_{AP}$, the number of PRBs assigned to each AP. Figure~\ref{fig:SFoutagePRB} plots the number of the users in outage normalized by the total number of the users for two user densities. All the users request the same data rate of 1.5Mbps. As shown in the figure, the outage decreases with $N_{AP}$ to a point where it is saturated such that further increase in $N_{AP}$ results in higher interference and hence, outage. In this example, $N_{AP} = 18$ gives the best performance for the given AP and user densities. In subsequent testing, we use a fixed value of $N_{AP} = 18$. This allows for a comparison of our results to the \emph{best-case scenario} for the fixed-allocation scheme.

\begin{figure}[h]
\center
\includegraphics [width=10cm,height=8cm,keepaspectratio]{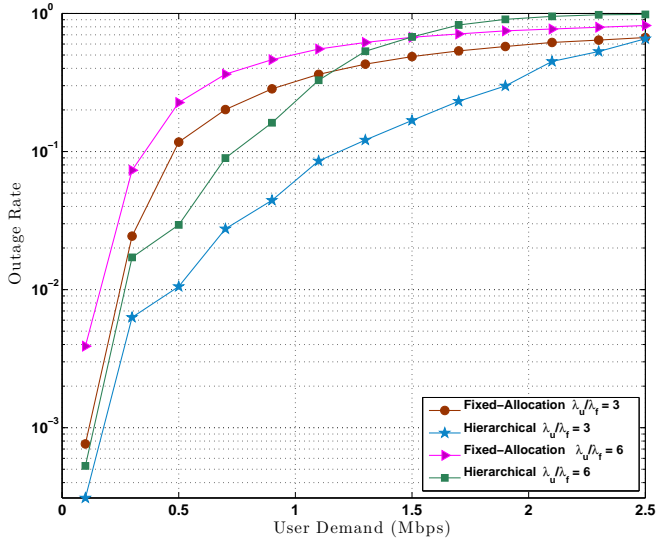}
\caption{Users at outage in both schemes versus the user demand. $\lambda_{f} = 1/(200m^{2})$, $\lambda_{u} = 3 \lambda_{f}$ and $\lambda_{u} = 6 \lambda_{f}$. $N_{AP} = 18$ for the fixed-allocation.}
\label{fig:OutageComp}
\end{figure}

\begin{figure}[htp]
\center
\includegraphics [width=10.5cm,height=8cm,keepaspectratio]{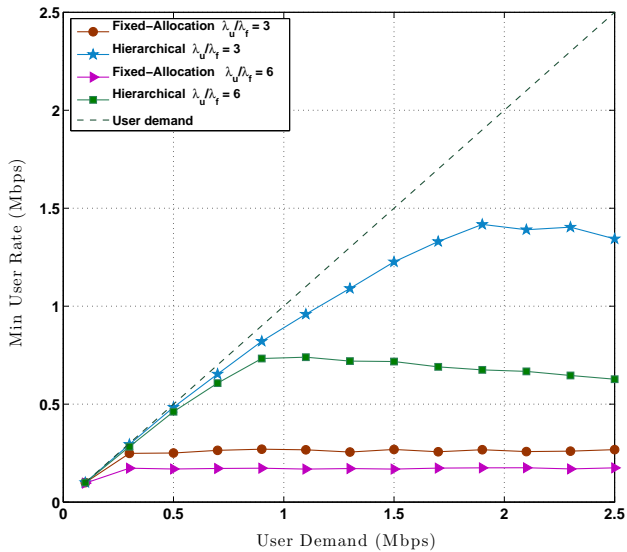}
\caption{Average minimum user achieved rate for both schemes versus the user demand. $\lambda_{f} = 1/(200m^{2})$, $\lambda_{u} = 3 \lambda_{f}$ and $\lambda_{u} = 6 \lambda_{f}$. $N_{AP} = 18$ for the fixed-allocation.}
\label{fig:MinRate}
\end{figure}

The outage (in a log-scale) for both schemes versus the user demand is shown in Fig.~\ref{fig:OutageComp}. As expected for both algorithms, the number of users in outage increases with the increase in the user demand. In both user-to-AP densities ($\lambda_{u}/\lambda_{f}$), there is an obvious gain with using the hierarchical scheme - the outage rate improves by up to an order of magnitude at the lower user demands. It is worth noting that at high user demands (above $R_k=$ 1.5Mbps for $\lambda_u/\lambda_f = 6$ and above $R_k = $ 2.5Mbps for $\lambda_u/\lambda_f =3$) the fixed-allocation scheme actually has a lower outage. This is to be expected since in the hierarchical scheme, any two APs that are less than $\tilde{r}$ meters apart are connected in the interference graph regardless of the degree of interference. This results in higher system load estimation and smaller number of PRBs allocated to \emph{each} AP in the system. In the fixed-allocation scheme on the other hand, due to the lack of any interference management, the effect of concurrent transmissions are added exactly according to the path loss model. Hence, the comparison here is  between the worst-case scenario of the hierarchical scheme and the best-case scenario of the fixed-allocation. Using weighted interference graphs and more sophisticated graph algorithms in Step 3 should improve the performance at the cost of increased computational complexity.

A significant advantage of the proposed scheme is to shift the available spectrum from the underloaded APs to the overloaded APs to achieve higher level of fairness over all the users in the network. Examining the minimum achieved user rate in the system in Fig.~\ref{fig:MinRate} shows that our proposed scheme achieves this goal. While both schemes converge to a constant value with the increase in the user demand, the hierarchical scheme reaches a higher value (more than twice the minimum rate in fixed-allocation) for both user densities. Any user achieved rate would fall between the minimum achieved rate and the user demand (the maximum rate assigned to the user represented by the dotted line). The closer the two are, the higher level of fairness is achieved. In other words, the hierarchical scheme achieves a higher degree of fairness and is more efficient in terms of allocating resources in comparison to the fixed resource allocation.

It is worth noting that this gain is higher in a system with a lower user-to-AP density. The reason is that in a network with independent user locations, the probability density function of the user distribution in a unit area (and hence the AP load) approaches a dampened normal distribution as opposed to a Poisson distribution with the increase in the user density. This effect corresponds to a smaller variance in the AP load in the system. The proposed scheme is most effective in systems with higher possibility of underloaded and overloaded APs existing at the same time which explains the higher gain in $\lambda_{u} = 3 \lambda_{f}$ compared to $\lambda_{u} = 6 \lambda_{f}$.

\begin{figure}
\center
\includegraphics[width=10cm,height=9cm,keepaspectratio]{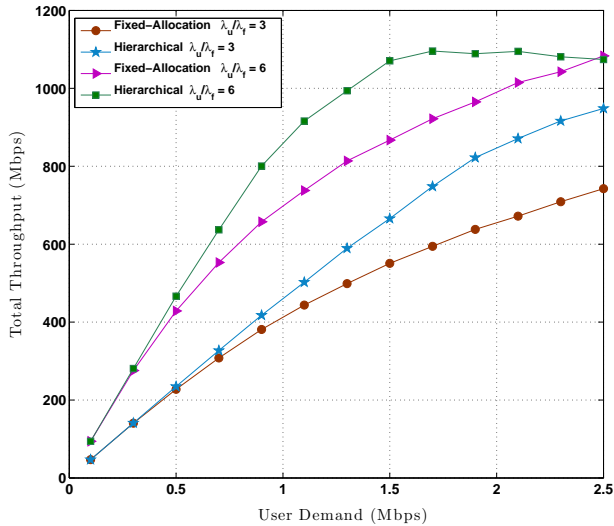}
\caption{Total throughput of the system for both schemes versus the user demand. $\lambda_{f} = 1/(200m^{2})$, $\lambda_{u} = 3 \lambda_{f}$ and $\lambda_{u} = 6\lambda_{f}$. $N_{AP} = 18$ for the fixed-allocation.}
\label{fig:Throughput}
\end{figure}

As a final comparison, Fig.~\ref{fig:Throughput} plots the total throughput of the system. The higher throughput in the hierarchical scheme is the result of higher user achieved rate as discussed above.

\section{Conclusion}\label{conclusion}
In this paper, we have proposed a hierarchical 4-stage resource allocation scheme for large-scale small-cell networks. The main advantage of the proposed scheme is decomposing a complex non-convex optimization problem into several smaller convex problems with smaller sets of optimization variables. The result is a low complexity scheme effective with a large problem size; in our simulations, the resource allocation can be achieved across as many as 314 APs.

The rationale behind the introduced hierarchy is as follows: user locations combined with various user demands result in a non-uniform distribution of the load in the system. Access points will experience very different load demands as shown in Fig.~\ref{fig:Nf_cdf2}. Hence, in an efficient allocation, resources should be dynamically allocated to meet this load. Here, this load is approximated at each AP by solving the related optimization problem based on the local information, i.e., users' demand and the average channel power. To do so, the APs do not require any global information. Load estimation and the last step of resource allocation at the APs would, in practice, be solved in parallel at each AP. Only a single step of graph coloring is executed at a central server - this centralized step ensures orthogonal allocations to APs that interfere with each other.

The central server only requires knowledge of the demands made by each AP which significantly reduces the signaling overhead. The results confirmed an increase in the user's achieved data rate with the proposed hierarchical scheme as apposed to the fixed resource allocation. Across a wide range of user rate demands, the scheme results in a significantly lower outage. Finally, the simulations confirm that the proposed scheme is most effective in systems with a higher possibility of underloaded and overloaded APs existing simultaneously.

\appendix[Complexity Analysis for Step 2: load estimation ]
The load estimation problem in~\eqref{prob2} is equivalent to finding the minimum of the following cost function
\begin{equation}
\begin{array}{ll}
\label{Lag}
\mathcal{L} = & \displaystyle \sum_{k \in S_{l}} n_{k}  + \sum_{k \in S_{l}} \mu_{k} \left [ R_{k} - n_{k}B\log_{2}\left ( 1 + \frac{P_{k} H_{k}}{n_{k}\sigma^{2}} \right) \right] \\
& + \mu_{0} \left(\sum_{k \in S_{l}} P_{k} - P_{tot} \right),
\end{array}
\end{equation}
where $\{\mu_{i}\}_{i = 0}^{|S_{l}|}$ are the Lagrangian multipliers. Differentiating~\eqref{Lag} with respect to $P_{k}$ and $n_{k}$, and setting each derivative to $0$, we obtain:
\begin{equation}
\label{Lag2}
\frac{\partial \mathcal{L}}{\partial n_{k}}  =  1 - \mu_{k} \mathcal{C}\left [ \ln \left ( 1 + \frac{P_{k} H_{k}}{n_{k}\sigma^{2}} \right)  - \frac{P_{k}H_{k}}{\sigma^{2} n_{k} \left(1 + \frac{P_{k} H_{k}}{n_{k}\sigma^{2}} \right)} \right] = 0, 
\end{equation}
\begin{equation}
\label{Lag3}
\frac{\partial \mathcal{L}}{\partial P_{k}}  =  - \frac{\mu_{k}\mathcal{C} \frac{H_{k}}{\sigma^{2}}}{1 + \frac{P_{k}H_{k}}{\sigma^{2} n_{k}}} + \mu_{0} = 0,
\end{equation}
for $k = 1, 2, \dots, M$, where $\mathcal{C} = B\log_{2}e$ and  $M = |S_{l}|$ are for simplicity of presentation. From~\eqref{Lag3}, we obtain:
\begin{equation}
\label{Lag4}
\frac{\mu_{k}H_{k}}{1 + \frac{P_{k}H_{k}}{n_{k}\sigma^{2}}} = \frac{\mu_{1}H_{1}}{1 + \frac{P_{1}H_{1}}{n_{1}\sigma^{2}}}.
\end{equation}
Combined with the power constraint $\sum_{k \in S_{l}} P_{k} = P_{tot}$, and the rate constraints $n_{k}B\log_{2}\left ( 1 + \frac{P_{k} H_{k}}{n_{k}\sigma^{2}} \right) = R_{k}$, $k = 1, 2, \dots, M$, there are $3M$ variables $\{P_{k}, n_{k}, \mu_{k}\}_{k=1}^{M}$ in the set of 3$M$ non-linear equations in~\eqref{Lag2}-\eqref{Lag4}. Iterative methods such as Newton-Raphson can be used to obtain the solution, with the complexity mainly due to finding the update direction.

Denote $\textbf{X} = [P_{1}, \dots, P_{M}, n_{1}, \dots, n_{M}, \mu_{1}, \dots, \mu_{M}]^{\intercal}$ as the variables and $\mathcal{\textbf{G}} = 0$ as the square system of non-linear equations. The update direction $\Delta \textbf{X}$ is found solving the following equation:
\begin{equation}
\mathcal{J}(\textbf{X})\Delta \textbf{X} = - \mathcal{\textbf{G}}(\textbf{X}),
\end{equation}
where $\mathcal{J}(\textbf{X})$ is the Jacobian matrix of $\mathcal{\textbf{G}}(\textbf{X})$ evaluated at $\textbf{X}$. Using Gauss-Jordan method, the complexity of the algorithm to solve for $\Delta \textbf{X}$ in each iteration is of the order $\mathcal{O}(M^{3})$. A special case is when equal transmit power is used on the subchannels; in this case, estimating the load at each AP has the complexity of the order $\mathcal{O}(M)$ (due to $M$ divisions at each AP).

\bibliographystyle{IEEEtran}
\bibliography{RefJournal}
\begin{IEEEbiography}[{\includegraphics[width=1in,
height=1.25in,clip,keepaspectratio]{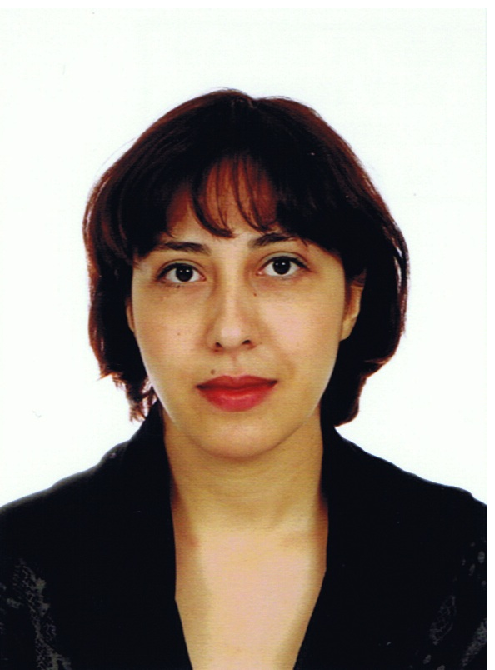}}]{Sanam Sadr}
(S'07) was born in Tehran, Iran. She received the B.Sc degree in electrical engineering from the University of Tehran, Iran in 2003 and the M.Sc. degree from Ryerson University, Toronto, Canada in 2007. She is currently a Ph.D student at the University of Toronto, Canada, where her research has focused on load balancing and resource allocation in heterogeneous networks using tools from stochastic geometry, point process theory, optimization and graph algorithms. She held a summer internship at Alcatel-Lucent Bell Labs, Stuttgart, Germany. She is the recipient of the Alexander Graham Bell Canada Graduate Scholarship-Doctoral in 2010. 
\end{IEEEbiography}

\vspace{-5in}
\begin{biography}[{\includegraphics[width=1in,
height=1.25in,clip,keepaspectratio]{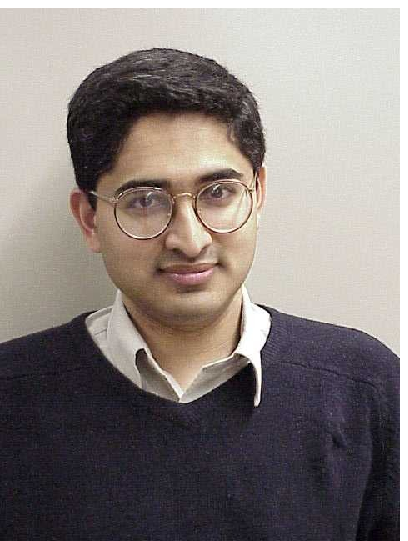}}]{Raviraj S. Adve}
(S'88, M'97, SM'06) was born in Bombay, India. He received
his B. Tech. in Electrical Engineering from IIT, Bombay, in 1990 and his
Ph.D. from Syracuse University in 1996. Between 1997 and August 2000, he
worked for Research Associates for Defense Conversion Inc. on contract with
the Air Force Research Laboratory at Rome, NY. He joined the faculty at the
University of Toronto in August 2000 where he is currently a Professor. Dr.
Adve's research interests include analysis and design techniques for
heterogeneous networks, energy harvesting networks and in signal processing
techniques for radar and sonar systems. He received the 2009 Fred Nathanson
Young Radar Engineer of the Year award.
\end{biography}

\end{document}